\documentclass[12pt]{article}
\usepackage{epsfig}
\usepackage{amssymb}
\usepackage{amsmath}
\usepackage{amsfonts}
\usepackage{graphicx}
\usepackage{mathrsfs}
\usepackage[dvips]{color}
\usepackage{multirow}

\newcommand{\bsigma}{\boldsymbol{\sigma}}

\newcommand{\R}{\mathbb{R}}
\newcommand{\C}{\mathbb{C}}
\newcommand{\Z}{\mathbb{Z}}

\newcommand{\fc}{\mathfrak{c}}

\newcommand{\ff}{\mathfrak{f}}

\newcommand{\fz}{\mathfrak{z}}

\newcommand{\fB}{\mathfrak{B}}

\newcommand{\fK}{\mathfrak{K}}
\newcommand{\bfK}{\boldsymbol{\mathfrak{K}}}

\newcommand{\bfe}{\mathbf{e}}

\newcommand{\bk}{\mathbf{k}}

\newcommand{\bp}{\mathbf{p}}
\newcommand{\bq}{\mathbf{q}}
\newcommand{\bfr}{\mathbf{r}}
\newcommand{\bx}{\mathbf{x}}
\newcommand{\by}{\mathbf{y}}

\newcommand{\bH}{\mathbf{H}}
\newcommand{\bI}{\mathbf{I}}

\newcommand{\bM}{\mathbf{M}}

\newcommand{\bU}{\mathbf{U}}

\newcommand{\cH}{\mathcal{H}}

\newcommand{\cF}{\mathcal{F}}

\newcommand{\cK}{\mathcal{K}}

\newcommand{\cM}{\mathcal{M}}

\newcommand{\cS}{\mathcal{S}}

\newcommand{\cU}{\mathcal{U}}

\newcommand{\be}{\begin{equation}}
\newcommand{\ee}{\end{equation}}
\newcommand{\bea}{\begin{eqnarray}}
\newcommand{\eea}{\end{eqnarray}}
\newcommand{\nn}{\nonumber}

\newcommand{\ed}{\end{document}}

\newcommand{\bi}{\begin{itemize}}
\newcommand{\ei}{\end{itemize}}

\newcommand{\bce}{\begin{center}}
\newcommand{\ece}{\end{center}}
\newcommand{\sA}{\mathscr{A}}
\newcommand{\sB}{\mathscr{B}}

\newcommand{\sD}{\mathscr{D}}

\newcommand{\sF}{\mathscr{F}}

\newcommand{\sT}{\mathscr{T}}
\newcommand{\sV}{\mathscr{V}}

\newcommand{\IM}{{\rm Im}}

\newcommand{\bPsi}{{\boldsymbol{\Psi}}}
\newcommand{\bPhi}{{\boldsymbol{\Phi}}}
\newcommand{\bPi}{{\boldsymbol{\Pi}}}
\newcommand{\bcK}{{\boldsymbol{\cK}}}
\newcommand{\bcM}{{\boldsymbol{\cM}}}

\newcommand{\bcH}{{\boldsymbol{\cH}}}
\newcommand{\bcU}{{\boldsymbol{\cU}}}

\newcommand{\bzero}{{\boldsymbol{0}}}

\newcommand{\for}{{\mbox{\rm for}}}

\begin{document}

\title{Transfer matrix formulation of stationary scattering in 2D and 3D: A concise review of recent developments}

\author{Farhang Loran\thanks{E-mail address: loran@iut.ac.ir}~ and
Ali~Mostafazadeh\thanks{E-mail address:
amostafazadeh@ku.edu.tr}\\[6pt]
$^{*}$Department of Physics, Isfahan University of Technology, \\ Isfahan 84156-83111, Iran\\[6pt]
$^\dagger$Departments of Mathematics and Physics, Ko\c{c}
University,\\  34450 Sar{\i}yer, Istanbul, Turkey}

\date{ }
\maketitle

\begin{abstract}

We review a recently developed transfer matrix formulation of the  stationary scattering in two and three dimensions where the transfer matrix is a linear operator acting in an infinite-dimensional function space. We discuss its utility in circumventing the ultraviolet divergences one encounters in solving the Lippman-Schwinger equation for delta-function potentials in two and three dimensions. We also use it to construct complex scattering potentials displaying perfect omnidirectional invisibility for frequencies below a freely preassigned cutoff.
\vspace{2mm}


\end{abstract}

\section{Introduction}

The introduction of the scattering (S) matrix and the discovery of the Lippman-Schwinger equation are among the most important achievements of the twentieth century theoretical physics. The former provides the basic ingredient of quantum scattering theory, while the latter serves as the main device for solving scattering problems. In one dimension, there is an alternative tool for performing scattering calculations, called the transfer matrix \cite{jones-1941,abeles,thompson} which, similarly to the S-matrix, stores the information about the scattering features of the system. The main advantage of the transfer matrix over the S-matrix is its composition property which makes it into an ideal tool for dealing with multilayer and locally periodic systems \cite{yeh,tjp-2020}.

Consider a short-range potential in one dimension, $v:\R\to\C$, so that $|v(x)|$ tends to zero faster than $|x|^{-1}$ as $x\to\pm\infty$. Then, every solution of the stationary Schr\"odinger equation,
    \be
    -\psi''(x)+v(x)\psi(x)=k^2\psi(x)~~~~x\in\R,
    \label{sh-eq-1D}
    \ee
satisfies
    \be
    \psi(x)\to A\pm e^{ikx}+B_\pm e^{-ikx}~~~\for~~~x\to\pm\infty,
    \label{asymp-1D}
    \ee
where $k\in\R^+$ is a wavenumber, and $A_\pm$ and $B_\pm$ are complex coefficients. The transfer matrix of the potential $v$ is a $2\times 2$ matrix $\bM$ that relates $A_\pm$ and $B_\pm$ according to
    \be
    \left[\begin{array}{c}
    A_+\\
    B_+\end{array}\right]=\bM
    \left[\begin{array}{c}
    A_-\\
    B_-\end{array}\right].
    \label{M-def-1D}
    \ee
This equation determines $\bM$ in a unique manner provided that it is independent of $A_-$ and $B_-$, \cite{epjp-2019}.

For $A_-=1$ and $B_+=0$ (respectively $A_-=0$ and $B_+=1$), $\psi(x)$ corresponds to a left-incident (resp.\ right-incident) wave, and the reflection and transmission amplitudes of the potential are respectively given by $R^l=B_-$ and $T^l=A_+$ (resp.\ $R^r=A_+$ and $T^l=B_-$). In view of these relations and (\ref{M-def-1D}), we can express $R^{l/r}$ and $T^{l/r}$ in terms of the entries $M_{ij}$ of $\bM$ according to $R^l=-M_{21}/M_{22}$, $R^r=M_{12}/M_{22}$ and $T^{l/r}=1/M_{22}$. Therefore, $\bM$ contains the complete information about the scattering properties of the potential.

Ref.~\cite{ap-2014} offers a dynamical formulation of stationary scattering (DFSS) in one dimension where the transfer matrix is identified with the S-matrix of an effective non-unitary two-level quantum system. Let $\psi$ be the general bounded solution of (\ref{sh-eq-1D}), and for each $x\in\R$, $\Psi_\pm(x):\R\to\C$ and $\bPsi(x):\R\to\C^{2\times 1}$ be the functions defined by
    \begin{align}
    &\big(\Psi_\pm(x))(p):=
    \frac{1}{2k}\,e^{\pm i kx}\,
    \left[k\psi(x)\pm i\,\psi'(x)\right],
    &&\bPsi(x):=\left[\begin{array}{c}
    \Psi_-(x)\\
    \Psi_+(x)\end{array}\right].
    \label{eq4}
    \end{align}
Then the stationary Schr\"odinger equation (\ref{sh-eq-1D}) is equivalent to the non-stationary Schr\"odinger equation, $i\partial_x\bPsi(x)={\bH}(x)\bPsi(x)$, where $x$ plays the role of ``time,''
${\bH}(x):=\frac{v(x)}{2k}e^{-ikx\bsigma_3}\bcK\,e^{ikx\bsigma_3}$ is an effective Hamiltonian, $\bsigma_3$ is the diagonal Pauli matrix, and $\bcK:=\left[\begin{array}{cc}1 & 1\\-1 & -1\end{array}\right]$.

The main reason for the introduction of the two-component state vector $\bPsi(x)$ is that, in view of \eqref{asymp-1D} and (\ref{eq4}), it satisfies $\bPsi(\pm \infty)=\left[\begin{array}{c}
    A_\pm\\
    B_\pm\end{array}\right]$.
This relation together and (\ref{M-def-1D}) allow us to express $\bM$ in terms of the evolution operator $\bU(x,x_0)$ for the Hamiltonian ${\bH}(x)$. Specifically, we have \(\bM=\bU(+\infty,-\infty)=\sT\exp\left[-i\int_{-\infty}^\infty dx\:{\bH}(x)\right]\), where $\sT$ denotes the time-ordering operation with $x$ playing the role of ``time.'' Note that the right-hand side of the preceding equation stands for the Dyson series expansion of $\bU(+\infty,-\infty)$. It offers a perturbative series expansion for $\bM$ which turns out to produce the exact solution of the scattering problem for single- and multi-delta function potentials in one dimension. Ref.~\cite{tjp-2020} provides a pedagogical review of this feature of DFSS and some of its notable applications, e.g., in constructing tunable unidirectional invisible potentials and single-mode inverse scattering.

Refs.~\cite{pra-2016,pra-2021} develop a higher-dimensional generalization of the DFSS where the role of the transfer matrix $\bM$ is played by a linear operator acting in an infinite-dimensional function space. This operator, which we also call the ``transfer matrix,'' has a canonical realization as a $2\times 2$ matrix $\widehat\bM$ with operator entries $\widehat M_{ij}$. Similarly to its one-dimensional analog, $\widehat\bM$ admits a Dyson series expansion and possesses a useful composition property.

One of the most remarkable benefits of DFSS in two and three dimensions is that its application to delta-function potentials in these dimensions \cite{pra-2016,pra-2021} circumvents the unwanted singularities of their standard treatments \cite{jackiw,manuel}. Among other applications of this approach to scattering theory are the discovery of a class of invisible (scattering-free) complex potentials \cite{pra-2021,prsa-2016,ol-2017,pra-2017,pra-2019} in two and three-dimensions and the construction of their electromagnetic counterparts \cite{jpa-2020}.

\section{Basic setup for stationary scattering in $D$ dimensions}

Consider the scattering problem defined by the stationary Schr\"odinger equation in $D+1$ dimensions,
    \be
    [-\partial_x^2-\nabla_\by^2+v(x,\by)]\psi(x,\by)=k^2\psi(x,\by),~~~~~~(x,\by)\in\R^{D+1},
    \label{sch-eq}
    \ee
where $(x,\by):=(x,y_1,y_2,\cdots,y_D)$ are Cartesian coordinates,
$v:\R^{D+1}\to\C$ is a scattering potential, $\psi:\R^{D+1}\to\C$ is a bounded function, and $\nabla_{\by}^2:=\sum_{j=1}^D\partial_{y_j}^2$. Suppose that we have chosen our coordinate system in such a way that the source of the incident wave and the detectors used to observe the scattered wave lie on the planes $x=\pm\infty$. If the source of the incident wave resides at $x=-\infty$ (resp.\ , $x=+\infty$) we speak of a left-incident (resp.\ right-incident) wave.

Let $\sF$ be the vector space of functions (tempered distributions) $f:\R^D\to\C$, $\cF_{\by,\bp}$ denote the Fourier transformation of a function of $\by$ evaluate at $\bp$, i.e.,
$\cF_{\by,\bp}\{f(\by)\}:=\int d^Dy\,e^{-i\bp\cdot \by} f(\by)$, and $\tilde f(\bp):=\cF_{\by,\bp}\{f(\by)\}$. Performing the Fourier transform of both sides of (\ref{sch-eq}) with respect to $\by$, we find
    \be
    -\tilde\psi''(x,\bp)+(\widehat\sV(x)\tilde\psi)(x,\bp)
    =\varpi(\bp)^2\,\tilde\psi(x,\bp),\quad\quad\quad (x,\bp)\in\R^{D+1},
    \label{sch-eq-FT}
    \ee
where  $\tilde \psi(x,\bp):=\cF_{\by,\bp}\{\psi(x,\by)\}$,
    \bea
    (\widehat\sV(x)\tilde f)(\bp)&:=&\cF_{\by,\bp}\{v(x,\by)f(\by)\}=
    \frac{1}{(2\pi)^D}\int d^Dq\,\tilde v(x,\bp-\bq)\tilde f(\bq),
    \label{v-def}\\
    \varpi(\bp)&:=&\left\{\begin{array}{ccc}
    \sqrt{k^2-\bp^2} & \for & |\bp|< k,\\
    i\sqrt{\bp^2-k^2} & \for & |\bp|\geq k,\end{array}\right.
    \label{varpi}
    \eea
and we have made use of the fact that $f(\by)=\cF^{-1}_{\bp,\by}\{\tilde f(\bp)\}:=\frac{1}{(2\pi)^D}\int d^Dp\, e^{i\bp\cdot\by}\tilde f(\bp)$.

An important feature of potential scattering in two and higher dimensions, which has no counterpart in one dimension, is the presence of evanescent waves. To elucidate their role, in the following, we confine our attention to the class of potentials $v$ that satisfy
    \be
    v(x,\by)=0~~\for~~x\notin[a_-,a_+],
    \label{supp-x}
    \ee
for some $a_\pm\in\R$ with $a_-<a_+$. Then, $\tilde v(x,\bp)=0$ for $x\notin[a_-,a_+]$, and  (\ref{sch-eq-FT}) gives  \([\partial_x^2+\varpi(\bp)^2]\tilde\psi(x,\bp)=0\) for \(x\notin[a_-,a_+].\)
Solving this equation and performing the inverse Fourier transform with respect to $\bp$, we can write $\psi$ in the form
    \be
    \psi=\psi_{\rm os}+\psi_{\rm ev},
    \label{decompose}
    \ee
where $\psi_{\rm os},\psi_{\rm ev}:\R^{D+1}\to\C$ are respectively functions representing the oscillating and evanescent waves outside the region defined by $a_-<x<a_+$ in $\R^{D+1}$, i.e.,    	\bea
    \psi_{\rm os}(x,\by)&=&\frac{1}{(2\pi)^D}
    \int_{\sD_k} \frac{d^Dp}{\varpi(\bp)}\;\left[A_\pm(\bp)\,e^{i\varpi(\bp)x}+B_\pm(\bp)\,e^{-i\varpi(\bp)x}\right]e^{i\bp\cdot\by}~~\for~~\pm x\geq \pm a_\pm,
    \label{psi-o1}\\
    \psi_{\rm ev}(x,\by)&=&\frac{1}{(2\pi)^D}
    \int_{\R^D\setminus \sD_k}\!\!\!\frac{d^Dp}{\varpi(\bp)}\: C_\pm(\bp)\,e^{\mp|\varpi(\bp)|x} e^{i\bp\cdot\by}~~\for~~\pm x\geq \pm a_\pm,
    \label{psi-e1}
    \eea
where \(\sD_k:=\left\{\bp|\bp^2\le k^2\right\}\), and $A_\pm,B_\pm,C_\pm\in\sF$ such that
    \begin{align}
    &A_\pm(\bp)=B_\pm(\bp)=0~~\for~~\bp\not\in\sD_k,
    &&C_\pm(\bp)=0~~\for~~\bp\in\sD_k.
    \label{ABC-bound}
    \end{align}
In particular, $A_\pm, B_\pm\in\sF_k$ where $\sF_k:\{\phi\in\sF|\phi(\bp)=0~\for~\bp\not\in\sD_k\}$. We also introduce,
    \begin{align}
    &\sB_-(\bp):=B_-(\bp)+C_-(\bp),    &\sA_+(\bp):=A_+(\bp)+C_+(\bp),
    \label{sA=}
    \end{align}
and employ (\ref{decompose}) -- (\ref{sA=}) to conclude that
    \bea
   \varpi(\bp) \tilde\psi(x,\bp)&=&\left\{
    \begin{array}{ccc}
    A_-(\bp)\,e^{i\varpi(\bp)x}+\sB_-(\bp)\,e^{-i\varpi(\bp)x} & \for & x\leq a_-,\\[6pt]
    \sA_+(\bp)\,e^{i\varpi(\bp)x}+B_+(\bp)\,e^{-i\varpi(\bp)x} & \for & x\geq a_+.
    \end{array}\right.
    \label{asym-1}
    \eea

The scattering solutions of the Schr\"odinger equation (\ref{sch-eq}) are particular bounded solutions of this equation that have the form $\psi(\bfr)=e^{i\bk_0\cdot\bfr}+\psi_{\rm scat}(\bfr)$, where $\bfr:=(x,\by)$ is the position vector, $\bk_0$ is the incident wave vector, $\psi_{\rm scat}$ signifies the scattered wave which satisfies
    \be
    \psi_{\rm scat}(\bfr)\to \left(ik^{-1}\right)^{\frac{2-D}{2}}r^{-\frac{D}{2}}\,e^{ikr}\ff(\hat{\bfr})    \quad\quad\for\quad\quad r:=|\bfr|\to\infty~{\rm and}~D=1,2,
    \label{scattering}
    \ee
$\ff(\hat{\bfr})$ is the scattering amplitude, and $\hat{\bfr}:=r^{-1}\bfr$. Let $\bp_0$ denote the projection of $\bk_0$ onto the orthogonal complement of the $x$-axis in $\R^{D+1}$. Then for a left-incident wave, $\bk_0=(\varpi(\bp_0),\bp_0)$,
	\begin{align}
	&A_-=(2\pi)^D\varpi(\bp_0)\delta_{\bp_0},
	\quad\quad \quad\quad \quad\quad B_+=0,
	\label{left}
	\end{align}
and as we show in \cite[Appendix~A]{pra-2016},
    \bea
    \ff(\hat{\bfr})&=&-\frac{i}{(2\pi)^{\frac{D}{2}}}\times\left\{
     \begin{array}{ccc}
     A_+(\bp)-(2\pi)^{D}\varpi( \bp_0)\delta(\bp-\bp_0)&\for&\hat\bx\cdot\hat\bfr>0\\
    B_-(\bp)&\for&\hat\bx\cdot\hat\bfr<0,\end{array}\right.
    \label{FL-3d}
    \eea
where $\delta_{\bp_0}(\bp):=\delta(\bp-\bp_0)$, $\bp$ is the projection of the scattered wave vector, $\bk:=k\hat\bfr$, onto the orthogonal complement of the $x$-axis, and $\hat\bx$ stands for the unit vector along the $x$-axis. In particular, $\bp=\bk-(\bk\cdot\hat\bx)\hat\bx=k[\hat\bfr-(\hat\bfr\cdot\hat\bx)\hat\bx]$ and $\varpi(\bp)=k|\hat\bfr\cdot\hat\bx|$. Similarly for a right-incident wave, $\bk_0=(-\varpi(\bp_0),\bp_0)$, and
	\begin{align}
	&A_-=0,
	\quad\quad\quad\quad \quad\quad  B_+=(2\pi)^D\varpi(\bp_0)\delta_{\bp_0},
	\label{right}\\
	&\ff(\hat{\bfr})=-\frac{i}{(2\pi)^{\frac{D}{2}}}\times\left\{
    \begin{array}{ccc}
    A_+(\bp)&\for&\hat\bx\cdot\hat\bfr>0,\\
    B_-(\bp)-(2\pi)^{D}\varpi( \bp_0)\delta(\bp-\bp_0)&\for&\hat\bx\cdot\hat\bfr<0.\end{array}\right.
    \label{FR-3d}
    \end{align}
According to (\ref{FL-3d}) and (\ref{FR-3d}), we can determine the scattering amplitude, i.e., solve the scattering problem, if we can express $B_-$ and $A_+$ in terms of $A_-$ and $B_+$. This is precisely what the S-matrix does, because according to (\ref{decompose}) -- (\ref{psi-e1}), for $x\to\pm\infty$, $\psi(\bfr)\to\psi_{\rm os}(\bfr)$, and $(B_-,A_+)$ and $(A_-,B_+)$ respectively determine the asymptotic ``out-going'' and ``in-going'' waves.

\section{Transfer matrix in higher dimensions}

Attempts at constructing and employing higher-dimensional generalizations of the transfer matrix has a long history \cite{pendry-1984,pendry-1996}. These were largely motivated by practical considerations. They involved a discretization of all but one of degrees of freedom and produced large numerical transfer matrices with a build-in composition property which allowed for numerical treatment of wave propagation and scattering. The developments reported in \cite{pra-2016,pra-2021} are of a completely different nature, for they introduce a fundamental notion of the transfer matrix which is amenable to analytic calculations.

By analogy to one dimension, we identify the transfer matrix in $D+1$ dimensions with the $2\times 2$ matrix $\widehat\bM$ with operator entries $\widehat M_{ij}:\sF_k\to\sF_k$ that satisfies
    \be
    \widehat{\bM}
    \left[\begin{array}{c}
     A_-\\
     B_-
     \end{array}\right]=
     \left[\begin{array}{c}
     A_+\\
     B_+\end{array}\right].
     \label{bM-def}
     \ee
In the following we use the term ``fundamental transfer matrix'' to refer to $\widehat\bM$. Similarly to the S-matrix, it relates $B_-$ and $A_+$ to $A_-$ and $B_+$; substituting (\ref{left}) and (\ref{right}) in (\ref{bM-def}), we find for left-incident waves:
	\begin{align}
	&\widehat M_{22}B_-=-(2\pi)^D\varpi(\bp_0)\widehat M_{21}\delta_{\bp_0},
	&&A_+=\widehat M_{12}B_-+(2\pi)^D\varpi(\bp_0)\widehat M_{11}\delta_{\bp_0},
	\label{left-eqs}
	\end{align}
and for right-incident waves:
	\begin{align}
	&\widehat M_{22}B_-=(2\pi)^D\varpi(\bp_0)\delta_{\bp_0},
	&&A_+=\widehat M_{12}B_-.
	\label{right-eqs}
	\end{align}
Eqs.~(\ref{FL-3d}), (\ref{FR-3d}), (\ref{left-eqs}), and (\ref{right-eqs}) reduce the solution of the scattering problem for the potential $v$ to the determination of $\widehat\bM$ and the solution of linear (integral) equations for $B_-$.

The fundamental transfer matrix enjoys a composition property similar to its one-dimensional analogue. The derivation of this property requires the use of a related object called the ``auxiliary transfer matrix'' \cite{pra-2021}. This is defined as the $2\times 2$ matrix $\widehat\bcM$ with operator entries $\widehat\cM_{ij}:\sF\to\sF$ fulfilling
	\be
    \widehat{\bcM}
    \left[\begin{array}{c}
     A_-\\
     \sB_-
     \end{array}\right]=
     \left[\begin{array}{c}
     \sA_+\\
     B_+\end{array}\right].
     \label{bcM-TM}
     \ee

Let, for each $x\in\R$, $\Phi_\pm(x):\R^D\to\C$ and $\bPhi(x):\R^D\to\C^{2\times 1}$ be the functions defined by
    \begin{align}
    &\big(\Phi_\pm(x)\big)(\bp):=
    \frac{1}{2}\,e^{\pm i \varpi(\bp)x}\,
    \left[\varpi(\bp)\tilde\psi(x,\bp)\pm i\,\tilde\psi'(x,\bp)\right],
    &&\bPhi(x):=\left[\begin{array}{c}
    \Phi_-(x)\\
    \Phi_+(x)\end{array}\right],
    \label{bPsi=}
    \end{align}
and introduce the effective Hamiltonian operator,
	\bea
    \widehat{\bcH}(x)&:=&\frac{1}{2}
    e^{-ix{\widehat\varpi}\bsigma_3}
    \widehat\sV(x)\,\bcK
    \, e^{i{\widehat\varpi}x\bsigma_3}{\widehat\varpi}^{-1},
    \label{bcH-def}
    \eea
where $\widehat\varpi:=\varpi(\widehat{\bp})$, and for every pair of functions $f,g:\R^D\to\C$, \(\big(f(\widehat{\bp})g\big)(\bp):=f(\bp)g(\bp)\).
Then the stationary Schr\"odinger equation (\ref{sch-eq}) is equivalent to
    \be
    i\partial_x \bPhi(x)=\widehat{\bcH}(x)\bPhi(x).
    \label{sch-eq-CH}
    \ee
Furthermore,  because $\widehat{\bcH}(x)=\widehat\bzero$ for $x\notin[a_-,a_+]$, where $\widehat\bzero$ is the zero operator acting in $\sF^{2\times 1}$, we can use (\ref{asym-1}), (\ref{bPsi=}), and (\ref{sch-eq-CH}) to conclude that
    \begin{align}
    &\lim_{x\to-\infty}\bPhi(x)=\bPhi(a_-)=
    \left[\begin{array}{c}
     A_-\\
     \sB_-\end{array}\right],
    &&\lim_{x\to+\infty}\bPhi(x)=\bPhi(a_+)=
    \left[\begin{array}{c}
     \sA_+\\
     B_+\end{array}\right].
     \label{bcM-TM-0}
    \end{align}
Eqs.~(\ref{bcM-TM}) and (\ref{bcM-TM-0}) imply
	\bea
    \widehat{\bcM}&=&\widehat\bcU(a_+,a_-)=
    \sT\exp\left[-i\int_{a_-}^{a_+} dx\:\widehat{\bcH}(x)\right]=\sT\exp\left[-i\int_{-\infty}^{\infty} dx\:\widehat{\bcH}(x)\right]
    \label{bcMb-def1}\\
    &=&\widehat\bI+\sum_{n=1}^\infty(-i)^n\int_{-\infty}^\infty\!\!dx_n\int_{-\infty}^{x_n}\!\!dx_{n-1}\cdots
    \int_{-\infty}^{x_2}\!\!dx_1\,\widehat{\bcH}(x_n)\widehat{\bcH}(x_{n-1})\cdots\widehat{\bcH}(x_1),
    \nn
    \eea
where $\widehat\bcU(x,x_0)$ is the evolution operator defined by the effective Hamiltonian (\ref{bcH-def}).

If $v(x,\by)$ vanishes for a range of values of $x$, $\widehat{\sV}(x)=\widehat 0$ and $\widehat{\bcH}(x)=\widehat\bzero$. This feature of $\widehat{\bcH}(x)$ implies the composition property of the (auxiliary) transfer matrix \cite{pra-2021}. As we explain in Ref.~\cite{pra-2021}, this follows from the semi-group property of the evolution operator. In order to benefit from the composition property of $\widehat\bcM$ in dealing with scattering problems, we explore the relationship between $\widehat\bcM$ and $\widehat\bM$.

Let us introduce,
	\begin{align}
    &\varpi_{\rm i}(\bp):=\IM[\varpi(\bp)]
    =\left\{\begin{array}{ccc}
    0 &\for&\bp\in\sD_k,\\
    \sqrt{\bp^2-k^2}&\for&\bp\not\in\sD_k,
    \end{array}\right.
    \label{omega-i}\\
    &\widehat\varpi_i:=\varpi_i(\widehat{\bp}),~~~~~~~~
    \bPsi(x):=e^{-{\widehat\varpi_i}\bsigma_3 x}\bPhi(x),~~~~~~~~
    \bPsi_\pm:=\left[\begin{array}{c}A_\pm\\ B_\pm\end{array}\right],
    \label{a1}\\
    &\widehat{\bH}(x)
    :=e^{-{\widehat\varpi_i}\bsigma_3 x}\widehat{\bcH}(x)\,
    e^{{\widehat\varpi_i}\bsigma_3 x}-i{\widehat\varpi_i}\bsigma_3,
    \label{bH-H}
    \end{align}
and \(\widehat\bPi_k:=\widehat\Pi_k\bI\), where \(\widehat\Pi_k:=\lim_{x\to\infty}e^{-\varpi_{\rm i}(\widehat{\bp})x}\) is the projection operator that maps $\sF$ onto $\sF_k$. Then, (\ref{bM-def}) reads
	\be
	\widehat{\bM }\bPsi_-=\bPsi_+,
	\label{P=MP}
	\ee
and we can use (\ref{sch-eq-CH}) to verify that $i\partial_x\bPsi(x)=\widehat{\bH}(x)\bPsi(x)$. Furthermore, \eqref{bcM-TM-0}, \eqref{omega-i}, and \eqref{a1} combined with \eqref{ABC-bound} and \eqref{sA=} give
    \begin{align}
    &\lim_{x\to\pm\infty}\bPsi(x)=\bPsi_\pm,
    &&{\widehat\varpi_i}\bPsi_\pm=0.
   \label{lim}
   \end{align}

Let $\bU(x,x_0)$ be the evolution operator for the Hamiltonian (\ref{bH-H}). Because \(\widehat{\bH}(x)=-i{\widehat\varpi_i}\bsigma_3\) for $x_0\leq x\leq a_-$ and $a_+\leq x_0\leq x$, we have $\bU(x,x_0)=e^{-(x-x_0){\widehat\varpi_i}\bsigma_3}$, and \eqref{bcMb-def1} and \eqref{lim} imply
    \begin{align}
    &\bPsi(x)=\lim_{x_0\to-\infty}\bU(x,x_0)\bPsi(x_0)=
    \lim_{x_0\to-\infty}e^{{\widehat\varpi_i}\bsigma_3x_0}\bPsi(x_0)=\widehat\bPi_k\bPsi_-=
    \bPsi_-\quad\quad \for \quad x\leq a_-,
    \label{Psi-}\\
    &\bPsi(x)=\bU(x,a_+)\bPsi_+=e^{-{\widehat\varpi_i}\bsigma_3 (x-a_+)}\bPsi_+=
    e^{-{\widehat\varpi_i}\bsigma_3 (x-a_+)}\widehat{\bcM}\bPsi_-,\quad\quad\mbox{for}\quad x\geq a_+.
    \label{a2}
    \end{align}
According to  \eqref{ABC-bound}, $\widehat\bPi_k\bPsi_-=\bPsi_-$, \(\widehat\bPi_ke^{-{\widehat\varpi_i}\bsigma_3 x}=\widehat\bPi_k\), and \(\widehat\bPi_k\bPsi(x)=\bPsi_+\) for $x>a_+$. We can use these relations together with \eqref{a2} to establish \(\bPsi_+=\widehat\bPi_k \widehat{\bcM}\,\widehat\bPi_k\bPsi_-\). Comparing this equation with (\ref{P=MP}), we arrive at
    \be
    \widehat{\bM}=\widehat\bPi_k \widehat{\bcM}\,\widehat\bPi_k.
    \label{comb}
    \ee

We have given the construction of the fundamental and auxiliary transfer matrices and derived some of their basic properties for potentials fulfilling (\ref{supp-x}). It is not difficult to see that these results extend to the class of potentials for which the solutions of the Schr\"odinger equation (\ref{sch-eq}) tend to plane waves for $x\to\pm\infty$. This is the case for short-range potentials which for $r\to\pm\infty$ tend to zero faster than $r^{-(D+1)}$, \cite{yafaev}.

\section{Perfect broadband invisibility}

The fundamental transfer matrix provides a convenient characterization of invisible potentials. According to (\ref{left}) -- (\ref{FR-3d}) and (\ref{bM-def}), omnidirectional invisibility for a wavenumber $k$ corresponds to the situation where \(\widehat{\bM}=\widehat\bI\) for this particular value of $k$ and arbitrary choices of the incident wave vector $\bk_0$ (arbitrary values of $\bp_0$).  Here and in what follows, we view $\widehat{\bM}$ as an operator acting in $\sF_k^{2\times 1}$, and use $  \widehat\bI$ to denote the identity operator for $\sF_k^{2\times 1}$.

\noindent {\bf Theorem~1}: Let $\alpha$ be a positive real number, $\hat\bfe$ be a unit vector that is perpendicular to the $x$-axis, and $v:\R^{D+1}\to\C$ be a short-range potential such that ${\tilde v}(x,\bfK)=0$ for $\bfK\cdot\hat\bfe\leq 2\alpha$. Then $v$ is omnidirectionally invisible for every wavenumber $k$ that does not exceed $\alpha$.

\noindent {\bf Proof}: In view of (\ref{bcMb-def1}) and (\ref{comb}), \(\widehat{\bM}=\widehat\bI\) holds, if for all $n\in\Z^+$ and $x_1,x_2,\cdots,x_n\in\R$,
    \be
    \widehat\bPi_k\widehat{\bcH}(x_n)
    \widehat{\bcH}(x_{n-1})\cdots
    \widehat{\bcH}(x_1)\widehat\bPi_k=\widehat\bzero.
    \label{condi-n}
    \ee
According to (\ref{bcH-def}), the entries of $\widehat{\bcH}(x)$ are given by
    $\widehat{\cH}_{jl}(x)=\frac{(-1)^{j+1}}{2}\:e^{i(-1)^j x{\widehat\varpi}}\widehat\sV(x)\,e^{i(-1)^{l+1} x{\widehat\varpi}}{\widehat\varpi}^{-1}$. Since functions of $\widehat{\bp}$ commute with the projection operator $\widehat\Pi_k$, this shows that (\ref{condi-n}) holds, if for all $f_1,f_2,\cdots, f_{n-1}\in\sF$,
    \be
    \widehat\Pi_k \widehat{\sV}(x_n)\,f_{n-1}(\widehat{\bp})
    \widehat{\sV}(x_{n-1})f_{n-2}(\widehat{\bp})
    \widehat{\sV}(x_{n-2})\cdots\cdots f_{1}(\widehat{\bp})
    \widehat{\sV}(x_1)\widehat\Pi_k=\widehat 0.
    \label{condi-n2}
    \ee
To establish this equation,  we introduce the function spaces
$\cS_\varsigma:=\{f\in\sF~|~f(\bp)=0~\for~p_\parallel\leq\varsigma\}$,
where $\varsigma\in\R$  and $p_\parallel:=\bp\cdot\hat\bfe$. Then for all $f_0\in\sF$, $\widehat\Pi_kf_0\in\cS_{-k}$. Next, let $\gamma\in\R$, $g\in\cS_\gamma$, and $\bp_\perp:=\bp-p_\parallel\hat\bfe$, so that we can use $(p_\parallel,\bp_\perp)$ to denote $\bp$. According to (\ref{v-def}),
    \bea
    \big(\widehat\sV(x)g\big)(\bp)&=&\frac{1}{(2\pi)^{D}}\int_{\gamma}^\infty dq_\parallel\:
    \int d^{D-1}\bq_\perp\tilde v(x,p_\parallel-q_\parallel,\bp_\perp-\bq_\perp)g(q_\parallel,\bq_\perp)\nn\\
    &=&\frac{1}{(2\pi)^{D}}\int_{-\infty}^{p_\parallel-\gamma} d\fK\int d^{D-1}\bq_\perp
    \tilde v(x,\fK,\bp_\perp-\bq_\perp)g(p_\parallel-\fK,\bq_\perp).
    \label{lemma1-3}
    \eea
By virtue of the hypothesis of the theorem, ${\tilde v}\in\cS_{2\alpha}$. Therefore \(\widehat\sV(x)\,g\in\cS_{2\alpha+\gamma}\).
This shows that $\widehat\sV(x)$ maps $\cS_\gamma$ to $\cS_{2\alpha+\gamma}$.
One easily verifies that
    \be
    \widehat{\sV}(x_n)\,f_{n-1}(\widehat{\bp})
    \widehat{\sV}(x_{n-1})f_{n-2}(\widehat{\bp})
    \widehat{\sV}(x_{n-2})\cdots\cdots f_{1}(\widehat{\bp})
    \widehat{\sV}(x_1)\Pi_k f_0(\bp)\in\cS_{2n\alpha-k}
    \label{eq43}
    \ee
For $k<\alpha$, $2n\alpha-k>k$, $\cS_{2n\alpha-k}\subset \cS_{k}$, and (\ref{eq43}) implies \eqref{condi-n2}.~$\square$\vspace{6pt}

The proof of Theorem~1 implies the following stronger result.\\[6pt]
\noindent {\bf Theorem~2}: Let $\xi\in[0,2\pi)$, $\alpha,\beta\in\R^+$, $k\in(0,\alpha]$, $\hat\bfe$ be a unit vector that is perpendicular to the $x$-axis, and $v:\R^{D+1}\to\C$ be a short-range potential such that for all $x\in\R$, ${\tilde v}(x,\bfK)=0$ for $\bfK\cdot\hat\bfe\leq\beta$. Then, $\widehat{\bM}=\widehat\bI$ for $\beta\geq 2\alpha$, and
    \be
    \widehat{\bM}=\widehat\bI+\sum_{n=1}^{\lceil 2\alpha/\beta-1\rceil} (-i)^n\!\!
            \int_{x_0}^x \!\!dx_n\int_{x_0}^{x_n} \!\!dx_{n-1}
            \cdots\int_{x_0}^{x_2} \!\!dx_1\,\widehat\bPi_k
            \widehat{\bcH}(x_n)\widehat{\bcH}(x_{n-1})\cdots
            \widehat{\bcH}(x_1)\widehat\bPi_k,
            \label{Thm-01-3D}
        \ee
for $0<\beta<2\alpha$, where $\lceil x\rceil$ stands for the smallest integer that is not smaller than $x$, and we assume that the operators appearing on both sides of (\ref{Thm-01-3D}) act in $\sF_k^{2\times 1}$.\\[6pt]
As shown in Ref.~\cite{pra-2021}, this theorem implies the following remarkable result on the discovery of potentials for which the first Born approximation in exact.\\[6pt]
\noindent {\bf Corollary}: Let $\alpha$, $\hat\bfe$, and $v$ be as in Theorem~2. Then the first Born approximation gives the exact expression for the scattering amplitude of $v$ for wavenumbers $k\leq\alpha$.

\section{Implicit regularization of delta-function potential in 2D}

The Lippmann-Schwinger equation for the delta-function potential,
	 \be
    v(\bfr)=\fz\,\delta^2(\bfr-\bfr_0),\ \ \ \ \ \bfr:=(x,y),
    \label{delta-1}
    \ee
with $\fz\in\C$ and $\bfr_0\in\R^2$, has the form
	\be
    \psi(\bfr)=e^{i\bk_0\cdot\bfr}+\fz\,\psi(\bfr_0)\,G(\bfr-\bfr_0),
    \label{b1}
    \ee
where $G(\bfr-\bfr_0)$ is the Green's function for the Helmholtz operator in two dimensions, i.e.,\linebreak \(\left(\nabla^2+k^2\right)G(\bfr-\bfr_0)= \delta^2(\bfr-\bfr_0)\),  that yields asymptotically out-going solutions of the Schr\"odinger equation (\ref{sch-eq}). It is well-known that
	\be
	G(\bfr)=\lim_{\epsilon\to^+0}\int\frac{d^2p}{(2\pi)^2}\frac{e^{i\bp\cdot\bfr}}{(k+i\epsilon)^2-\bp^2}=-\frac{i}{4}H_0^{(1)}(kr),
	\label{G=}
	\ee
where  \(H_0^{(1)}\) stands for the zero-order Hankel function of the first kind. For $r\to\infty$, $G(\bfr-\bfr_0)\to
-\sqrt{i/8\pi kr}\,e^{-ik\bfr_0\cdot\hat\bfr}e^{ikr}$. In view of this relation and Eqs.~(\ref{scattering}) and (\ref{b1}), the scattering amplitude of the potential (\ref{delta-1}) takes the form
	\be
	\ff(\hat\bfr)=-\frac{\fz\,\psi(\bfr_0)}{2\sqrt{2\pi}}\, e^{-ik\bfr_0\cdot\hat\bfr} .
	\label{f=delta}
	\ee
	
We can determine $\fz\,\psi(\bfr_0)$ by setting $\bfr=\bfr_0$ in (\ref{b1}). This gives $\fz\,\psi(\bfr_0)=e^{i\bk_0\cdot\bfr_0}/[\fz^{-1}-G(\bzero)]$. The problem with this calculation is that $G(\bzero)$ is logarithmically divergent. Therefore, setting $\bfr=\bfr_0$ in (\ref{b1}) is forbidden. One can however attempt to regularize $G(\bfr)$ and employ a coupling constant renormalization to remove its singularity \cite{jackiw}.

Let $G_\Lambda(\bfr)$ be the regularized Green's function obtained by restricting the domain of the integral in (\ref{G=}) to a sphere of radius $\Lambda$, so that $G(\bfr)=\lim_{\Lambda\to\infty}G_\Lambda(\bfr)$. Then, it is easy to show that $G_\Lambda(\mathbf{0})= -\frac{1}{4\pi}\ln\left(\frac{\Lambda^2}{k^2}-1\right)-\frac{i}{4}$. Substituting $G_\Lambda(\bfr)$ for $G(\bfr)$ in (\ref{b1}) and repeating the above calculation of $\fz\,\psi(\bfr_0)$, we find
	\be
	\fz\,\psi(\bfr_0)=\frac{e^{i\bk_0\cdot\bfr_0}}{\fz^{-1}-G_\Lambda(\bzero)}=
	\frac{e^{i\bk_0\cdot\bfr_0}}{\tilde\fz(k)^{-1}+\frac{i}{4}},
	\label{z-renormalized}
	\ee
where $\tilde\fz(k)$ is the renormalized coupling constant given by
$\tilde\fz(k):=\left[\fz^{-1}+\frac{1}{4\pi}\ln\left(\frac{\Lambda^2}{k^2}-1\right)\right]^{-1}$. Demanding the latter not to depend on $\Lambda$, we can express it in the form, ${\tilde\fz(k)}=\left[{\tilde\fz(k_{\rm ref})}^{-1}-\frac{1}{2\pi}\ln\left(\frac{k}{k_{\rm ref}}\right)\right]^{-1}$,
where $k_{\rm ref}$ is a reference wavenumber. Substituting (\ref{z-renormalized}) in (\ref{f=delta}), we obtain
	\begin{align}
	&\ff(\hat\bfr)=-\sqrt{\frac{2}{\pi}}\,\frac{e^{-i\bfr_0\cdot(k\hat\bfr-\bk_0)}}{4\,\tilde\fz(k)^{-1}+i}.
    	\label{LSch}
	\end{align}
	
Next, we examine the application of our transfer-matrix formulation of stationary scattering to the delta-function potential (\ref{delta-1}) for right-incident waves. The same analysis applies to the left-incident waves.

Let $(a,b)$ be the coordinates of $\bfr_0$, so that $v(x,y)=\fz\,\delta(x-a)\delta(y-b)$. To determine the fundamental transfer matrix for this potential, we substitute \eqref{delta-1} in (\ref{v-def}) and use the resulting equation in \eqref{bcH-def} to show that $\widehat\bcH(x)=
\frac{\fz}{2}\delta(x-a) e^{-ia\widehat\varpi\bsigma_3}\widehat\sV_b\,\bcK\,
e^{ia\widehat\varpi\bsigma_3}\widehat\varpi^{-1}$, where $\big(\widehat\sV_b g\big)(p):=
e^{-ibp}\cF^{-1}_{q,b}\{g(q)\}$. Because \(\bcK^2=\mathbf{0}\), $\widehat\bcH(x_1)\widehat\bcH(x_2)=\bzero$. This makes the Dyson series expansion of the auxiliary transfer matrix \eqref{bcMb-def1} terminate, and we can use \eqref{comb} to show that
$
	\widehat{\bM}=
	e^{-ia\widehat\varpi\sigma_3}\left(\widehat\bPi_k-\frac{i\fz}{2}\,\bcK\,		
	\widehat\bPi_k \widehat\sV_b\, \widehat\bPi_k\widehat\varpi^{-1}\right)e^{ia\widehat\varpi\sigma_3}.
$
Next, we read off the entries $\widehat M_{12}$ and $\widehat M_{22}$ of $\widehat{\bM}$ from this equation and use them to express (\ref{right-eqs}) as		
	\begin{align}
	&\fB_-=-\frac{i\fz}{2}\widehat\bPi_k \widehat\sV_b\, \widehat\bPi_k\widehat\varpi^{-1}\fB_-+
	2\pi\varpi(p_0)e^{-ia\varpi(p_0)}\delta_{p_0},
	\label{B-eq}\\
	&A_+= e^{-2ia\widehat\varpi}B_--2\pi\varpi(\bp_0)e^{-2ia\varpi(p_0)}\delta_{p_0},
	\label{A-eq}
	\end{align}
where $\fB_-:=e^{-ia\widehat\varpi}B_-$. It is not difficult to see that $\big(\widehat\bPi_k \sV_b\, \widehat\bPi_k\widehat\varpi^{-1}\fB_-\big)(p)=\fc\,e^{-ibp}\chi_k(p)$, where $\fc:=\cF_{q,b}^{-1}\{\varpi(q)^{-1}\fB_-(q)\}$, $\chi_k(p):=1$ for $|p|<k$, and $\chi_k(p):=0$ for $|p|\geq k$. Inserting this relation in (\ref{B-eq}), and using the right-hand side of the resulting equation to compute $\fc$, we find a linear equation whose solution is $\fc=(1+i\fz/4)^{-1}e^{i(-a\varpi(p_0)+bp_0)}=(1+i\fz/4)^{-1}e^{i\bk_0\cdot\bfr_0}$. Here we have also made use of the identity $\int_{-k}^k dq/\sqrt{k^2-q^2}=\pi$. In view of the formula we obtained for $\fc$, Eqs.~(\ref{B-eq}) and (\ref{A-eq}), and the fact that $B_-=e^{ia\widehat\varpi}\fB_-$, we have
$A_+(p)=A_0\chi_k(p) e^{-i(a\varpi(p)+bp)}$ and $B_-(p)=A_0\chi_k(p)e^{-i(-a\varpi(p)+bp)}+2\pi\varpi(p_0)\delta(p-p_0)$, where $A_0:=-2ie^{i\bk_0\cdot\bfr_0}/(4\fz^{-1}+i)$. Substituting these equations in (\ref{FR-3d}), we arrive at $\ff(\hat\bfr)=-\sqrt{2/\pi}\,e^{-i(\bk-\bk_0)\cdot\bfr_0}/(4\,\fz^{-1}+i)$. Because $\bk=k\hat{\bfr}$, this equation is in perfect agreement with (\ref{LSch}) provided that we interpret $\fz$ as the physical coupling constant (which in the standard treatment of the problem is identified with $\tilde\fz(k)$). Notice the application of the transfer matrix to the delta-function potential (\ref{delta-1}) we never encounter singularities and there is no need for the renormalization of the coupling constant $\fz$; the transfer-matrix formulation of stationary scattering has a build-in regularization feature in 2D. The same holds in 3D \cite{pra-2021}.\vspace{12pt}

\noindent{\bf Acknowledgements}:
This work has been supported by the Scientific and Technological Research Council of Turkey in the framework of the project 120F061 and by Turkish Academy of Sciences.

\ed